\documentclass[12pt]{iopart}

\usepackage{xcolor}
\usepackage{ulem}
\usepackage{soul}
\usepackage{hyperref}

\begin{document}

\title[Quantum researcher mobility]{Quantum researcher mobility: the wonderful wizard of Oz who paid for Dorothy’s Visa fees}
\author{Mehul Malik}
\address{Institute of Photonics and Quantum Sciences (IPAQS), Heriot-Watt University, Edinburgh, United Kingdom}
\ead{m.malik@hw.ac.uk}
\author{Elizabeth Agudelo}
\address{Atominstitut, Technische Universit\"at Wien, 1020 Vienna, Austria}
\address{Institute for Quantum Optics and Quantum Information - IQOQI Vienna, Austrian Academy of Sciences, Boltzmanngasse 3, 1090 Vienna, Austria}
\ead{elizabeth.agudelo@tuwien.ac.at}
\author{Ravi Kunjwal}
\address{Centre for Quantum Information and Communication, Ecole polytechnique de Bruxelles, CP 165, Universit\'e libre de Bruxelles, 1050 Brussels, Belgium}
\ead{ravi.kunjwal@ulb.be}
\vspace{10pt}

\begin{abstract}
Historically, science has benefited greatly through the mobility of researchers, whether it has been
due to large-scale conflict, the search for new opportunities or a lack thereof. 
Today’s world of strict global immigration policies, exacerbated by the COVID-19 pandemic, places inordinate hurdles on the mobility of all researchers, let alone quantum ones. 
Exorbitant visa fees, the difficulty of navigating a foreign immigration system, lack of support for researchers’ families, and explicit government policy targeting selected groups of immigrants are all examples of things that have severely impacted the ability of quantum researchers to cross both physical and scientific borders. 
Here we clearly identify some key problems affecting quantum researcher mobility and discuss examples of good practice on the governmental, institutional, and societal level that have helped, or might help, overcome these hurdles. 
The adoption of such practices worldwide can ensure that quantum scientists can reach their fullest potential, irrespective of where they were born.
\end{abstract}


It would be na\"ive to say that science knows no borders. 
While it is certainly true that science aspires to
transcend geographic and societal borders, the knowledge we create today and how we do it is very much a reflection of the world we live in. 
Scientific funding systems closely mirror geopolitical structures around the world, subsequently leading to a large disparity in access to funding and, as a result, the scale of research efforts in countries around the world. 
In contrast to this, one could argue that the scientific researcher really knows no borders. 
The pursuit of science involves the exchange and creation of knowledge in a manner that is ideally unconstrained by political and religious beliefs, or socioeconomic and geographic means. 
The field of contemporary quantum information science aspires, as it should, to this ideal. 
From South America to Asia, quantum scientists today are spread all over the globe, and quantum science relies heavily on international collaboration and the exchange of ideas across many of these boundaries.
The quantum world is replete with examples of leading figures in the field who emigrated from their countries of origin at an early career stage, as well as prominent scientists who returned after stints abroad to establish a quantum research programme in their home countries. 
Take the example of Pakistani-American physicist Nergis Mavalvala, who moved to the US at the age of eighteen to pursue a bachelor’s degree in Physics and Astronomy. Today, Nergis is a long-standing member of the LIGO team that first detected gravitational waves and herself has made pioneering contributions to the field of quantum measurement science. 
She is also Professor of Astrophysics and Dean of the School of Science at MIT, and is often viewed as a role model for aspiring female scientists of South Asian origin \cite{Venkatraman2012}. 
On the flip side, leading figures such as Jian-Wei Pan and Luiz Davidovich are examples of quantum scientists who obtained their PhDs in a foreign country (Austria and the US, respectively) and subsequently returned to their countries to establish world-leading research programs in quantum information science. 
Jian-Wei Pan led the launch of the first quantum satellite Micius that first distributed entanglement to Earth from space and is leading China’s recent quantum efforts \cite{MITPan}.
Luiz Davidovich has made major contributions to the fields of decoherence, entanglement, laser theory, and quantum metrology and has played a fundamental role in the development of experimental quantum optics in South America \cite{Davidovich}. 
On a broader scale, recent studies have shown that nearly a quarter of Nobel laureates in Physics received the award for work that they had done outside their country of origin. 
In addition, researchers who move around the world can be seen to have greater scientific impact than their non-mobile counterparts \cite{Petersen2018,Gewin2018}.
If you can afford moving where the opportunities and the funding abound, your career will clearly benefit.

It is clear from the examples and studies referenced above that researcher mobility not only provides significant benefits for advancing science as a whole, but also scientific research efforts in the destination country as well as the researcher’s country of origin. 
Central to these examples was the ability of these scientists to travel to a foreign country unhindered, and assimilate within the local scientific and social culture in order to develop their skills and research careers further. 
It is perhaps not surprising that all these examples involve countries of the Global North \cite{wiki:gn_gs}, in Europe and North America. 
Science as practiced today is heavily centred around the Global North. 
As such, the immigration policies of countries in these parts of the world have a disproportionate effect on the careers of academic researchers from across the world. 
Indeed, mobility between citizens of countries in the Global North is relatively easy, at least in comparison with the hurdles faced by researchers from the Global South when looking for the very same opportunities. 
In this article, we draw upon our own experiences as researchers from the Global South (Asia, Latin America), having worked in the Global North (the United Kingdom, Canada, and Europe), to highlight what we believe are the most serious problems affecting researcher mobility. 
We also discuss examples of good practice at the governmental, institutional, and societal level that have helped, or might
help, remove barriers to the international mobility of scientific researchers looking for such opportunities.

\section*{Exorbitant visa fees, non-eligibility for reimbursement}
Most researchers looking to work in a country different from their own need to obtain permission to work, i.e., a work permit or visa from the government of that country. 
Enter the problem of visa fees. 
The amount a researcher has to pay for a work visa varies wildly from country to country. For example, the visa fees for a family of four to move to the United Kingdom (UK) for five years are approximately 17 500 GBP \cite{UKVisaFees}, while in Austria the same family would pay around 600 EUR for a two-year permit \cite{AustriaVisaFees}. 
One might wonder how the average scientific researcher can afford such exorbitant, indeed exploitative, visa fees as in the UK. 
The answer is quite simply that they cannot. 
A potential result of such practices is a disproportionately larger distribution of talented researchers in countries where visa fees are affordable or waived. In parallel, access to opportunities offered by mobility are restricted to a privileged few \cite{lse}. 
The example of the UK here sticks out like a sore thumb—the very high fees are disguised as a “healthcare surcharge” for funding the UK’s National Health Service (NHS). 
However, almost all researchers are employees of UK institutions and hence pay mandatory national insurance, which also funds the NHS. 
Hostile policies such as this form of double taxation place a significant burden on researchers and their families, who are already facing significant costs associated with moving to a new country. 
To make things worse, researchers on a temporary visa are also excluded from social benefits in the UK, which has a direct impact on the well-being of their dependents.

Aside from visa fees for taking up work, even a Standard Visitor Visa in the UK for an academic visit can put an undue burden on the finances of researchers. 
Here's a concrete example: to attend a week-long conference in the UK, an Indian citizen would have to pay 190 GBP for an academic visitor visa for up to six months that does not allow for paid work.\footnote{A Standard Visitor Visa for tourism, on the other hand, costs 95 GBP.} 
On the other hand, a European citizen can make the same visit without any visa. 
This extra cost could still be mitigated if one could charge it to one’s research grant, but to make matters worse, many scientific funding agencies specifically exclude visa fees as a reimbursable expense when travelling for work, while allowing for expenses related to transportation and accommodation (e.g., the Chargé de Recherche fellowship of the Belgian Fonds de la Recherche
Scientifique-FNRS). 
At the same time, even if funding agencies did reimburse visa fees, there would still be a penalty for having the ‘wrong’ passport: the Indian researcher would have an effectively smaller research budget than a European researcher on the same fellowship or grant, even if personal costs are mitigated by allowing reimbursement of visa fees. 
All this, again, is quite separate from the extra mental burden of having to reapply for a visa and the associated paperwork for every visit outside the six-month window \cite{lse} or the additional financial cost of applying for a longer-term Standard Visitor visa (the two, five, and ten-year options cost 361, 655, and 822 GBP, respectively). 
A non-EU scientist looking to visit the Schengen area for a conference or an academic visit faces the same difficulties, albeit with the Visa fee waived \cite{SchengenVisaFees}.

\section*{Short-term contracts, visa renewals, funding agency rules}
Even after successfully navigating the visa system and having worked in a new country for a number of years, every international researcher is faced with the inevitable hurdle of visa renewal. 
The seemingly straightforward process (additional money aside) of renewing one’s right to work is drastically complicated
by the way many academic jobs are structured. 
Postdoctoral contracts can range from a few months to a few years, and are often reliant on fragmented funding, i.e., multiple short-term contracts associated with different funders or research institutes rather than one long-term contract. 
A work permit or visa is normally tied to a specific employer, so a simple change of contract can trigger the need for a visa renewal.
The renewal process itself can take several months, during which time one is normally not allowed to start work with the new employer and hence is effectively unemployed. 
There are numerous stories of researchers who have had to leave their country of work during this renewal process, incurring further costs from cancelled rental contracts and airfare to and from their home country—all due to an inflexible and
dysfunctional immigration system.

\section*{Advantage home students/researchers and mental health ramifications}
One has to consider that in a highly competitive field such as quantum science, inordinate obstacles such as these tilt the playing field disproportionately in favour of students and researchers with no immigration restrictions. 
For example, a postdoc from the European Union (EU) can legally reside in any European country without a work contract, allowing them to ‘easily’ navigate the complex world of short-term academic contracts (a process that is difficult enough without immigration restrictions). 
Additionally, professors or research supervisors looking to hire new PhD students or postdocs might prefer candidates with an uncomplicated immigration situation, simply because funding for the position may lapse by the time the visa or work permit arrives.
Critical steps one would normally take for building a successful career in research—developing an independent research programme, applying for competitive fellowships and grants, networking at scientific workshops—all suffer heavily under the cloud of uncertainty surrounding one’s basic ability to continue living in the same place. 
The mental health toll of living with such uncertainty on an individual cannot be discounted, and can result in many talented international scientists leaving their field of research, being forced to return home to further uncertainty, or worse \cite{Dolan_case}.
Institutions should also be wary of ignoring the systemic nature of the problem and deflecting attention by treating it as a mental health issue, for example by recommending that a student or employee stressed due to their immigration situation see a mental health professional rather than trying to address the root cause.
While access to mental health support is critical in such situations, it cannot serve as a substitute for actual change in policies that lead to, or exacerbate, mental health crises.

\section*{Cultural hurdles to mobility}
Aside from governmental and institutional hurdles to researchers’ mobility, there are cultural hurdles one faces when moving to a new country to live and work, however temporary. 
These hurdles are an inevitable consequence of the globalisation of science and the movement of researchers across geographic borders. 
We would like to highlight two specific examples of such hurdles. 
First, the emotional cost of having to relocate and re-establish a home base in a new city or institution every time one changes a fixed-term contract can be significant. 
Second, the wide adoption of English as the language of science benefits native English speakers disproportionately and can take a toll on the opportunities that are accessible to scientists who are not native English speakers.

\subsection*{Importance of a local support network}
A fulfilling scientific career rests on more than just the science \cite{Gewin_2019}. 
Scientists need the emotional safety net of friendship and family as much as anyone else. 
Mobility, while justified professionally, can result in alienation from one’s family, separation from loved ones, or the lack of a reliable network of friends in a new culture. 
Culture shocks and the lack of belonging one might feel typically has an unavoidable impact on one’s productivity and well-being. For someone who may have been part of a close-knit family unit, such separation can cause extreme emotional distress. 
It can be difficult to build a sense of belonging from scratch with each move and the shorter the fixed-term contract is, the more difficult it gets to establish a trustworthy community of friends and colleagues that one can count on in times of need. 
For many researchers, moving is not even an option for these or other reasons. 
For example, scientists with disabilities may have special needs that are not catered to their satisfaction (or at all) in every place they might want to move to, even if they are somehow willing to risk not having the network of friends and family that keeps them productive and healthy in their place of work. 
The personal choice of leaving everything behind and embarking on new adventures that come with migrating as a scientist could perhaps be more easily made if there were a better and more consistent strategy from governments and institutions to help with moving one’s family, and funding agencies were more open to and flexible about the development of research in multiple places, including allowing remote working arrangements when feasible. 
Remote working during the COVID-19 pandemic has shown that this is certainly possible if funding agencies and research institutions are willing to work out the necessary modalities.

\subsection*{Language hegemony in scientific communication -- the Latin American case}
Having a common language is fundamental for the exchange of scientific communication. 
English happens to be the pre-dominant language of such communication. 
A scientist must be proficient in English to attend English-language conferences, read and write English-language papers, and have English-language discussions. 
If they want to have influential and globally recognized publications, these must be in English, since the most prestigious scientific journals are published in this language \cite{Huttner-Koros2015}. 
There are clear disadvantages that come from this language barrier which preserves the global gap in science. 
Let us consider the Latin American case \cite{Valenzuela-Toro2021}. 
English is usually not taught in many schools in the region \cite{Amano2016}. 
Therefore, most researchers there learn it when, as undergraduate students, they realize that it is necessary in order to be able to study books and articles for advanced courses.
This prevents them from accessing many scientific books, blogs, videos, or podcasts long before their undergraduate studies, as the best material is typically available in English. 
Of course, it is worth mentioning that this is not the case for all scientists in the region since there is a strong correlation between English proficiency and higher socioeconomic origin. 
In any case, once a researcher reaches a reasonable comfort-level with English, they are welcome to the scientific sphere, or at least that is what we would like to believe. 
However, even after writing and submitting articles in English, researchers whose first language is not English often have to deal with bias and their work is often discarded without review. 
Indeed, many good results may never be submitted to English language journals. 
Between the lack of language resources and high teaching loads, professors often have no means to go through all the work it takes to go from a good result to a publishable
draft in English. 
As a consequence, such results may be ignored since non-English articles, when they get written at all, may reach a very limited audience.

\section*{Examples of good practice - suggested solutions}
There are a few examples of countries around the world with immigration systems that attempt to address some of the problems we have outlined above. 
As an example of good governmental practice, Austria has a specific visa category for researchers (‘Forscher’), where the (relatively low compared to the UK) application fees for the main applicant are normally waived. 
In addition, immigrants are eligible for social benefits, allowing them to access the same quality of life for themselves and their families as that provided to Austrian citizens. 
As another example of good practice, the UK Global Talent visa category allows researchers to change employers without reapplying for a visa, but is only open to applicants who can demonstrate that they meet specific definitions of ‘talent’ and involves the same exorbitant visa fees.
However, it does provide some form of stability for researchers who are bound to short-term contracts or have no choice but to change employers often. 
On the level of funding agencies, the European Research Council allows PIs some flexibility for reimbursing immigration costs, enabling them to level the hiring process. 
As an institutional example, IQOQI Vienna provides new employees with a welcome booklet that includes critical information about health insurance, visa procedures, and funded language immersion courses, while clearly outlining its organisational structure.
These are but a few drops in the large bucket of possible solutions and practices that governments and institutions can adopt to help and encourage researcher mobility. 
We summarise these examples and some more in the list below.\\ 

\noindent Governments can:
\begin{itemize}
    \item Establish a visa category for researchers and their families with reduced or waived visa fees.
    \item Establish a similar low-cost or free visa category for short-term research visits such as conferences and internships.
    \item Allow researchers working in a country and their families to access national social benefits.
    \item Allow researchers to stay in the country while their visa renewal is being processed.
    \item Not tie a researcher visa to a specific employer or institution, allowing for freedom of movement within the same country.
    \item Regulate the processing time of visas in agreement with funding agencies’ conditions.
\end{itemize}    
    
\noindent Principal investigators can:
\begin{itemize}
     \item Liaise with funding agencies and their institutions to ensure that research grants can be used for reimbursing visa and mobility costs.
    \item Take additional mobility costs into account and be sensitive to the emotional toll of an international move when hiring foreign researchers and students.
    \item Facilitate short Bachelors/Masters student visits from the global south, funding for which is often covered by European grants.
\end{itemize}

\noindent Institutions/organisations can:
\begin{itemize}
    \item Have dedicated and trained staff member(s) to assist with immigration issues and liaise with government officials when necessary.
    \item  Have a pool of common funding earmarked for meeting visa costs for researchers who might incur them on scientific visits abroad. At the very least, funding agencies should reimburse visa costs as a legitimate travel expense for work.
    \item  Establish interest-free loan programmes for exorbitant visa costs, or reimburse them outright, as is currently being done by many academic institutions in the UK \cite{Edinburgh,HWU}.
    \item Provide early access to mobility assistance funds or at the very least, an advance salary payment, prior to their physical move.
    \item  Provide early access to mobility assistance funds or at the very least, an advance salary payment, prior to their physical move.
    \item  Provide language immersion courses with funding covered by institutions, funding agencies, or government bodies.
    \item  Organise conferences with clear guidelines to address unconscious bias in the review process, such as that associated with geographic origin \cite{Q-Turn}.
    \item Give researchers attending international conferences ample time to apply for visas and provide financial support when possible.
    \item Encourage multilingual events, perhaps using technology to generate automatic captions or sign language support in online talks and seminars.
    \item  Provide support for the families of researchers, for example by compiling information about local international support groups, childcare, education, language and cultural immersion courses.
\end{itemize}

\section*{Conclusion}
The foreign researcher’s experience is not far from that of Dorothy in the Wonderful Wizard of Oz \cite{Oz}---having one’s entire home uprooted and thrown into a new land, meeting interesting (and sometimes strange) new colleagues, navigating a winding career path replete with complex bureaucracy and grant applications, and negotiating with powerful and mysterious benefactors. 
Now imagine if Dorothy also faced the additional obstacles presented by a modern immigration system. 
It is clear that foreign quantum researchers have a lot to offer the new countries they choose to call home, as well as their own home countries, should they choose to move back. 
In this article, we have tried to summarise the main hurdles placed on the mobility of foreign researchers, and outlined concrete solutions that governments, institutions, and principal investigators can follow in order to help make this process easier. This is by no means an exhaustive list, and we invite you to be a part of this important and open conversation. 
For example, we have not discussed the difficulties of building a pension in a typical academic career that involves jumping from country to country. 
It is our hope that this article will lead to changes in policy that help reduce the unnecessary burden placed on foreign researchers by complex and prohibitive immigration systems, and allow them to focus purely on the advancement of quantum science.

\section*{Acknowledgements}
We would like to acknowledge the many unnamed people (administrators, colleagues, and friends) who have sacrificed their time to help us navigate the complex web of immigration policies and hurdles that we have encountered in our own careers. 
MM acknowledges support by the UK Engineering and Physical Sciences Research Council (EPSRC) (EP/P024114/1) and the European Research Council (ERC) Starting Grant PIQUaNT (950402). 
EA acknowledges funding from the Austrian Science Fund (FWF): M3151. 
RK is supported by the Chargé de Recherche fellowship of the Fonds de la Recherche Scientifique—FNRS, Belgium.

\section*{Data availability statement}
No new data were created or analysed in this study.

\section*{ORCID iDs}

Mehul Malik \href{https://orcid.org/0000-0001-8395-160X}{https://orcid.org/0000-0001-8395-160X}\\
Elizabeth Agudelo \href{https://orcid.org/0000-0002-5604-9407}{https://orcid.org/0000-0002-5604-9407} \\
Ravi Kunjwal \href{https://orcid.org/0000-0002-3978-5971}{https://orcid.org/0000-0002-3978-5971}

\section*{Bibliography}
\bibliographystyle{unsrt}
\bibliography{bibfile}

\begin{thebibliography}{10}

\bibitem{Venkatraman2012}
Vijaysree Venkatraman.
\newblock \href{https://doi.org/10.1126/science.caredit.a1200061}{Gravitational
  wave researcher succeeds by being herself}.
\newblock {\em Science careers}, 2012.

\bibitem{MITPan}
{Martin Giles}.
\newblock
  \href{https://www.technologyreview.com/2018/12/19/1571/the-man-turning-china-into-a-quantum-superpower/}{The
  man turning China into a quantum superpower}.
\newblock {\em MIT Technology Review}, 2018.

\bibitem{Davidovich}
{Luiz Davidovich - Interview}.
\newblock
  \href{https://twas.org/article/luiz-davidovich-you-must-go-home-again}{Luiz
  Davidovich: You must go home again}, 2014.

\bibitem{Petersen2018}
Alexander~M. Petersen.
\newblock \href{http://doi.org/10.1098/rsif.2018.0580}{Multiscale impact of
  researcher mobility}.
\newblock {\em J. R. Soc. Interface.}, 15:20180580, 2018.

\bibitem{Gewin2018}
{Virginia Gewin}.
\newblock \href{https://doi.org/10.1038/d41586-018-07499-3}{Why you should move
  country}.
\newblock {\em Nature Career News}, 2018.

\bibitem{wiki:gn_gs}
{Global North and Global South}.
\newblock
  \href{https://en.wikipedia.org/wiki/Global_North_and_Global_South}{Global
  North and Global South --- {W}ikipedia{,} The Free Encyclopedia}, 2021.

\bibitem{UKVisaFees}
{United Kingdom Visa Fees}.
\newblock
  \href{https://www.gov.uk/government/publications/visa-regulations-revised-table/home-office-immigration-and-nationality-fees-6-april-2022
  }{UK Home Office Immigration and Nationality Fees: 6 April 2022}, 2022.

\bibitem{AustriaVisaFees}
{Austrian Researcher Permit Fees}.
\newblock
  \href{https://oead.at/en/to-austria/entry-and-residence/settlement-permit-researcher
  }{OeAD - Austrian Agency for Education and Internationalisation} - settlement
  permit researcher, 2022.

\bibitem{lse}
{Bathsheba Okwenje}.
\newblock
  \href{https://blogs.lse.ac.uk/impactofsocialsciences/2019/07/19/visa-applications-emotional-tax-and-privileged-passports/}{Visa
  applications: emotional tax and privileged passports}, 2019.

\bibitem{SchengenVisaFees}
{Short Stay Schengen Visa Fees, Netherlands}.
\newblock
  \href{https://www.netherlandsworldwide.nl/countries/united-states/travel/applying-for-a-short-stay-schengen-visa}{Applying
  for a short-stay Schengen visa in the United} states - netherlands worldwide,
  2022.

\bibitem{Dolan_case}
{Andrew Grant}.
\newblock
  \href{https://physicstoday.scitation.org/do/10.1063/pt.6.2.20170803a/full/}{A
  paper on field theory delivers a wake-up call to academics}, 2017.

\bibitem{Gewin_2019}
{Virginia Gewin}.
\newblock \href{https://www.nature.com/articles/d41586-019-00902-7}{Five
  scientists explain how they decided whether to move to another country for}
  their work or studies, 2019.

\bibitem{Huttner-Koros2015}
{Adam Huttner-Koros}.
\newblock
  \href{https://www.theatlantic.com/science/archive/2015/08/english-universal-language-science-research/400919/}{The
  Hidden Bias of Science’s Universal Language}, 2015.

\bibitem{Valenzuela-Toro2021}
Ana~M. Valenzuela-Toro and Mariana Viglino.
\newblock \href{https://www.nature.com/articles/d41586-021-02601-8}{How Latin
  American researchers suffer in science}, 2021.

\bibitem{Amano2016}
{Tatsuya Amano}.
\newblock
  \href{https://www.cam.ac.uk/research/news/languages-still-a-major-barrier-to-global-science-new-research-finds}{Languages
  still a major barrier to global science, new research finds}, 2016.

\bibitem{Edinburgh}
{Immigration Fee Assistance Programmes}.
\newblock
  \href{https://www.ed.ac.uk/human-resources/international-staff/international-staff/after-1-january-2021/immigration-fee-assistance
  }{University of Edinburgh - Human Resources}, 2022.

\bibitem{HWU}
{Interest-free loan scheme for staff working in the UK}.
\newblock
  \href{https://www.hw.ac.uk/uk/services/docs/hr/policies/immigration-guidance-visa-fees-interest-free-loan-scheme.pdf
  }{Heriot-Watt University - Human Resources}, 2022.

\bibitem{Q-Turn}
{Ana Bel\'{e}n} Sainz.
\newblock \href{https://arxiv.org/abs/2202.06867}{Q-Turn: Changing Paradigms In
  Quantum Science}.
\newblock {\em arXiv}, page 2202.06867, 2022.

\bibitem{Oz}
L.~Frank Baum.
\newblock {\em The Wonderful Wizard of Oz}.
\newblock George M. Hill Company, Chicago, IL, 1900.

\end{thebibliography}

\end{document}